\documentclass[twocolumn,tighten]{aastex63}

\usepackage{times,natbib,graphicx,amsmath,multirow,xspace}
\usepackage{xcolor}
\usepackage{lineno}
\usepackage{rotating}
\usepackage{longtable}

\newcommand{\nustar}{NuSTAR\xspace}

\newcommand{\nicer}{NICER\xspace}

\newcommand{\fluxcgs}{ergs~cm$^{-2}$~s$^{-1}$\xspace}

\newcommand{\rin}{$R_{\rm in}$\xspace}
\newcommand{\rg}{$R_{g}$\xspace}
\newcommand{\risco}{$R_{\mathrm{ISCO}}$\xspace}

\newcommand{\relxill}{{\sc relxill}\xspace}
\newcommand{\relxillns}{{\sc relxillNS}\xspace}

\newcommand{\xillver}{{\sc xillver}\xspace}
\newcommand{\xillverco}{{\sc xillverCO}\xspace}
\newcommand{\xillverns}{{\sc xillverNS}\xspace}
\newcommand{\source}{\mbox{4U~0614$+$091}\xspace}

\newcommand{\ox}{\ion{O}{8}\xspace}

\definecolor{pink}{RGB}{212,76,133}
\definecolor{sky}{RGB}{51,160,232}
\definecolor{lavender}{RGB}{170,120,240}

\begin{document}

\title{A View of the Long-Term Spectral Behavior of Ultra Compact X-Ray Binary 4U 0614+091}

\correspondingauthor{D. L. Moutard}
\email{moutard@umich.edu}

\author[0000-0003-1463-8702]{D.~L.~Moutard}
\affiliation{Department of Physics \& Astronomy, Wayne State University, 666 West Hancock Street, Detroit, MI 48201, USA}
\affiliation{Department of Astronomy, University of Michigan, 1085 S. University, Ann Arbor, MI 48109, USA}
\author[0000-0002-8961-939X]{R.~M.~Ludlam}
\author[0000-0002-8294-9281]{E.~M.~Cackett}
\affiliation{Department of Physics \& Astronomy, Wayne State University, 666 West Hancock Street, Detroit, MI 48201, USA}
\author[0000-0003-3828-2448]{J.~A.~Garc\'{i}a}
\affiliation{Cahill Center for Astronomy and Astrophysics, California Institute of Technology, 1200 E. California Blvd, MC 290-17, Pasadena, CA, 91125, USA}
\author[0000-0003-2869-7682]{J.~M.~Miller}
\affiliation{Department of Astronomy, University of Michigan, 1085 S. University, Ann Arbor, MI 48109, USA}
\author[0000-0002-4794-5998]{D.~R.~Wilkins}
\affiliation{Kavli Institute for Particle Astrophysics and Cosmology, Stanford University, 452 Lomita Mall,Stanford, CA 94305-4085, USA}

\begin{abstract}
In this study, we examine 51 archival \nicer observations and 6 archival \nustar observations of the neutron star (NS) ultra-compact X-ray binary (UCXB) \source, which span over 5 years. The source displays persistent reflection features, so we use a reflection model designed for UCXBs, with overabundant carbon and oxygen (\xillverco) to study how various components of the system vary over time. The flux of this source is known to vary quasi-periodically on a timescale of a few days, so we study how the various model components change as the overall flux varies. The flux of most components scales linearly with the overall flux, while the power law, representing coronal emission, is anti-correlated as expected. This is consistent with previous studies of the source. We also find that during observations of the high-soft state, the disk emissivity profile as a function of radius becomes steeper. We interpret this as the corona receding to be closer to the compact object during these states, at which point the assumed power law illumination of \xillverco may be inadequate to describe the illumination of the disk.\\
\end{abstract}

\section{Introduction} \label{sec:intro}
In low-mass X-ray binary (LMXB) systems, a neutron star (NS) or black hole (BH) typically interacts with a main sequence star $\lesssim 1 M_\odot$. This interaction takes place via Roche-lobe overflow, where the stellar companion fills its Roche lobe, until the material begins to fall towards the NS or BH, and eventually circularizes into an accretion disk. Typical LMXBs have orbital periods on the timescales of hours or days, but some display a much shorter orbital period \citep{avakyan23}. This subclass of LMXB is called an ultra-compact X-ray binary (UCXB) and is generally comprised of a NS or BH, and a H-poor degenerate companion like a white dwarf (WD) or He star. The orbital period required to be considered a UCXB is typically $<$ 80 minutes, which would not be possible for larger main sequence or H-rich stars without directly colliding with the compact accretor \citep{savonije86,bahramian23}. Because UCXB companion stars are chemically distinct from a main sequence star, the abundances in the accretion disk will also differ quite greatly from those LMXBs which are not ultra-compact. These systems typically have negligible amounts of hydrogen and overabundance of oxygen \citep{nelemans04,nelemans06}. Aside from being useful avenues of studying accretion physics and probing the nature of compact objects in a unique way, these sources will also produce gravitational waves on the order of mHz. This will be detectable by future multi-messenger missions such as LISA \citep{chen20}, so understanding and classifying these systems well in advance is useful. 

LMXB systems are typically defined as having a few primary components. These are a disk which produces thermal X-rays \citep{shakura73}, a corona close to the compact object which produces hard, non-thermal X-rays via the Compton scattering of disk photons, and in the case of NS systems, thermal photons from the NS itself \citep{syunyaev91}. These contribute to the overall continuum of the X-ray spectrum, but coronal photons will also illuminate the disk and be reprocessed by the material therein. This leaves a signature as the photons are reprocessed, often in the form a relativistically broadened iron K-shell fluorescence emission around 6-7 keV\citep{fabian89}. In the case of UCXBs, the overabundance of oxygen means that this feature may be screened as interactions occur more readily with more abundant elements in the disk, and instead a more dominant \ox feature appears in the softest X-rays, at around 0.67 keV \citep{koliopanos13,koliopanos21}. These additional features are considered part of the reflection spectrum, and are believed to come from the innermost region of the disk \citep{tanaka95,bhattacharyya07}. Based on the broadening of these lines we can determine several useful quantities such as the location of the inner disk with respect to the compact object and the inclination of the system. 

One such UCXB is \source. This source was discovered in 1978 by the Uhuru mission and considered to be a binary system \citep{swank78}. Later observations of type-I X-ray bursts determined that the compact object was a neutron star \citep{brandt92}, followed by later spectral studies verifying the presence of elements that make the companion likely to be WD \citep{juett01,nelemans04}. The orbital period was measured by \cite{shahbaz08} using optical data to be approximately 50 minutes. This was done by directly observing modulations in the optical light curve caused by the uneven heating of the donor star.

\cite{moutard23} used reflection modeling to study \source as the flux changed. The reflection model used therein is \xillverco, which is a modification of the \xillver reflection table \citep{garcia13}. The key difference is that \xillverco has abundances designed for CO WD, with reduced amounts of hydrogen and helium \citep{madej14}. The source appears to have a quasi-periodic flux, varying on the timescale of $\sim$a week. The study concluded that the flux of most of the spectral components are correlated with the overall flux, but found that in the lowest flux state, the non-thermal flux was at its {\it highest}, and the disk appeared to truncate to about twice the innermost stable circular orbit (\risco). \risco is the closest a particle can stably orbit a compact object before falling onto it, and is equal to 6 \rg (\rg $= GM/c^2$, with 6\rg = 12.4 km for a non-rotating $1.4 M_\odot$ NS). A parallel is drawn to BH LMXB systems, where typically at lower luminosities, the disk truncates \citep{done07}. The trend is less clear in NS systems, as the complicated magnetic fields affect the location of the inner disk, but since both NSs and BHs are compact accretors, this comparison is not unwarranted. 

In this paper, we extend the study from \cite{moutard23}, supplementing the 4 \nicer observations with additional archival observations. Our goal is to study the long term behaviors of the system as the overall flux varies, and to test further analogies between NS and BH accretion systems. While the previous study benefited from simultaneous \nicer and \nustar observations, the observations only cover roughly 1 period of the quasi-periodic flux oscillations. Because of this, it is difficult to draw conclusions about the system more generally. By extending the study we are able to draw more robust conclusions about model parameters and their relation to the variation in flux over time. This will also serve as a test of the limits of \xillverco. The model currently assumes the bulk of the illumination is from the non-thermal corona, but it is well understood that in higher luminosity states, the continuum shifts to be dominated by softer, thermal X-rays \citep{lin07}. We choose \nicer because of the large number of archival observations. \nicer also provides excellent coverage in the band containing the \ox feature, which is key to understanding UCXBs. We also supplement this with 6 archival \nustar observations. \nustar can be useful for constraining the overall continuum shape, as the band extends into the hard X-rays. Although there are many observations of \source using Swift, the source is quite bright. Therefore, the observations are taken in windowed mode to account for additional photon pileup. These pileup effects can introduce additional uncertainties to the emission spectrum of a source \citep{miller10,ng10}, so we opt to use \nicer and \nustar, which do not suffer from such pileup effects. These instruments also allow us to directly compare to our previously published literature. Section \ref{sec:obs} of this paper discusses the details of the archival observations and the selection criteria, Section \ref{sec:results} provides details on the modeling and reports the relevant results of fitting spectra to these models, and Section \ref{sec:disc} discusses the physical interpretations of these results. The final section, Section \ref{sec:conc}, summarizes this work.

\section{Observations and Data Reduction} \label{sec:obs}
\begin{figure}[h!t]
\begin{center}
\includegraphics[width=0.48\textwidth, trim = 0 0 0 0, clip]{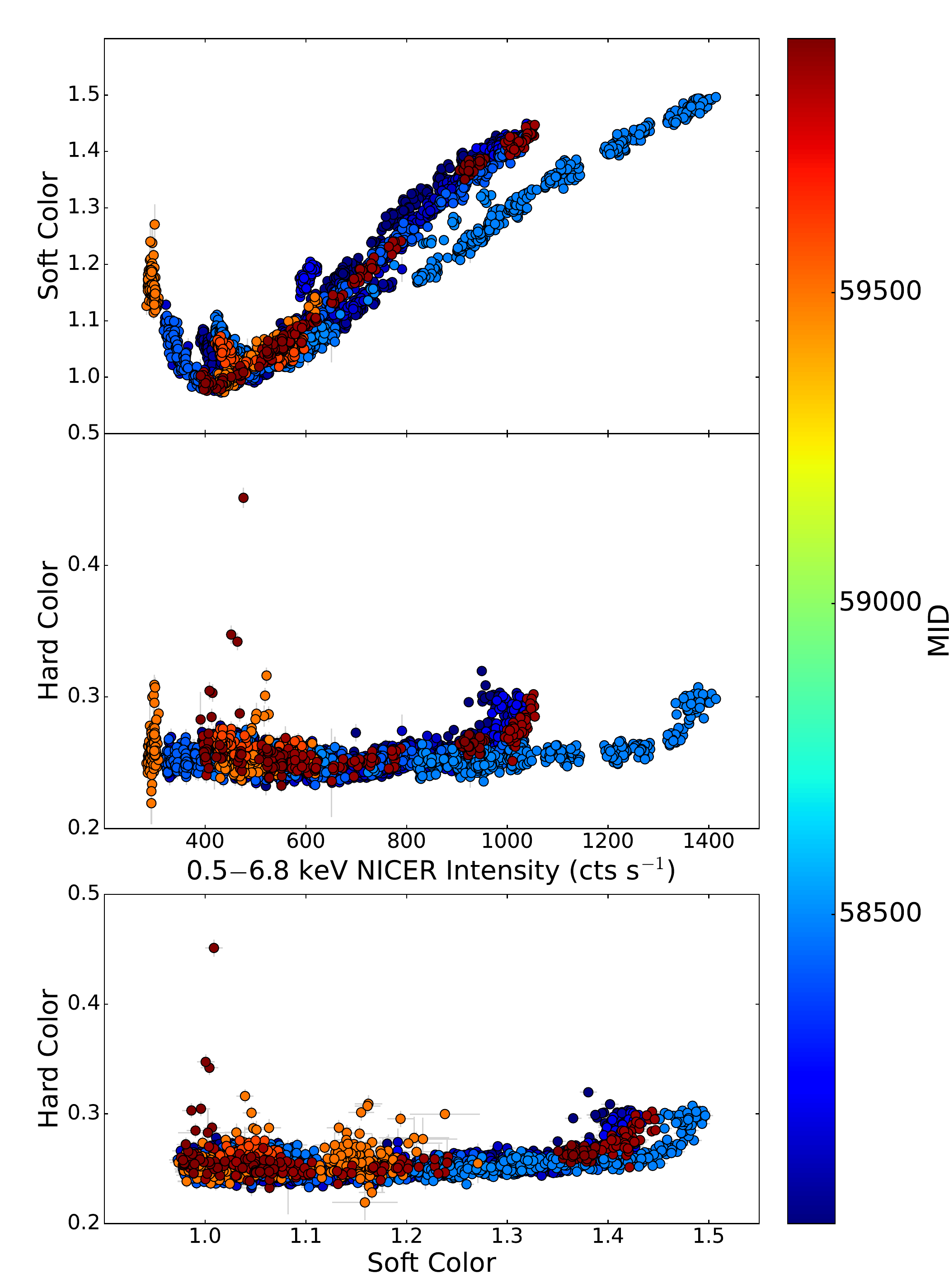}
\caption{The hardness intensity diagram for the soft color (1.1-2.0 keV/0.5-1.1 keV) and the hard color (3.8-6.8 keV/2.0-3.8 keV), as well as the color-color diagram for all 51 observations. The observations are binned into 150s segments for clarity. We see that the hard color stays relatively constant throughout all observations, but the soft color varies more distinctly over time.}
\label{fig:cid}
\end{center}
\end{figure}

\begin{figure*}[h!t]
\begin{center}
\includegraphics[width=0.98\textwidth, trim = 0 9 0 5, clip]{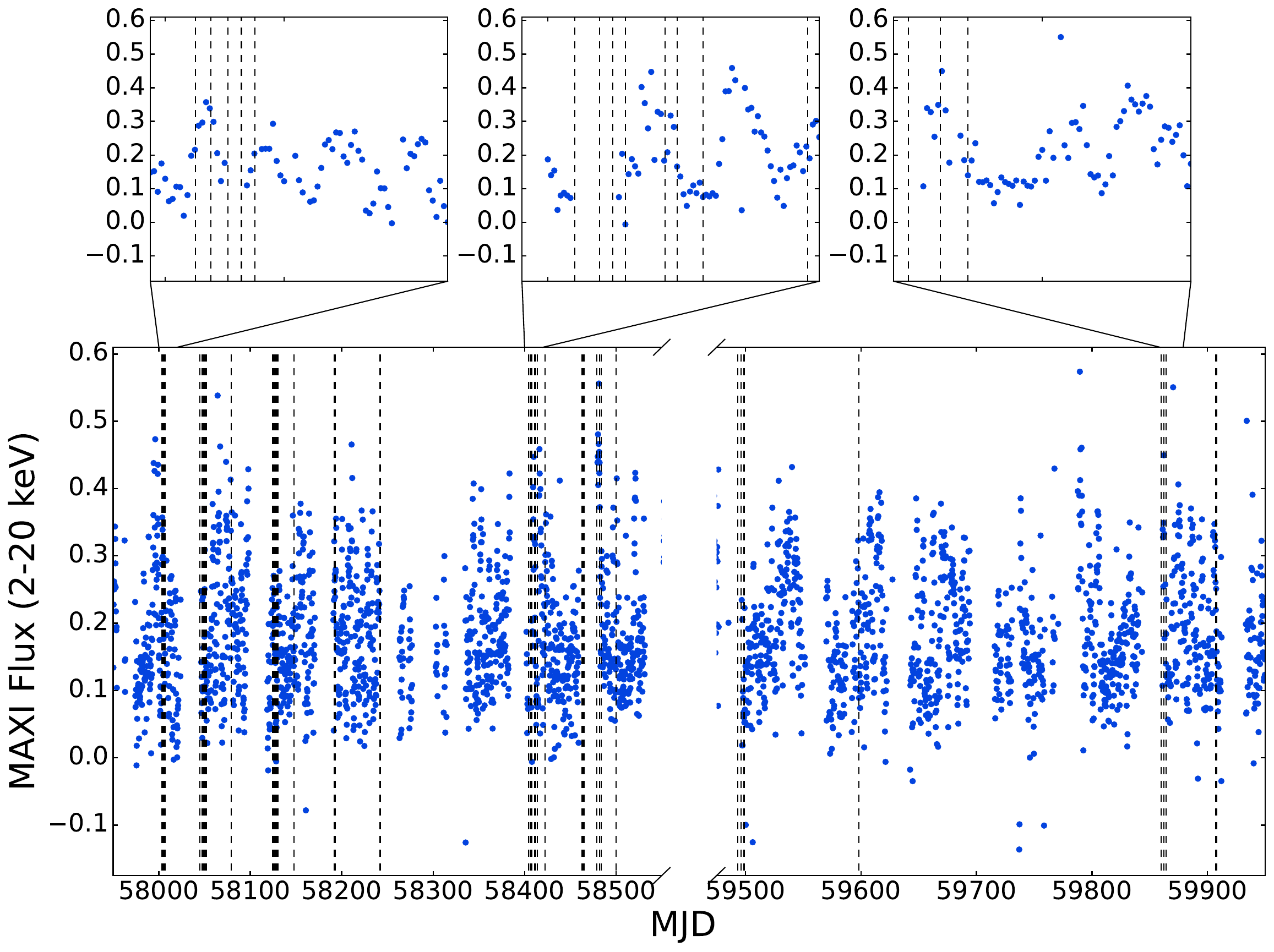}
\caption{Long term MAXI light curve in blue points with the start date of each observation overlaid as dashed lines. The top row of plots highlights specific 20-day regions of the light curve, demonstrating that the source is captured in various states. These zoomed-in plots also display the quasi-periodic nature of the flux of the source.}
\label{fig:maxi}
\end{center}
\end{figure*}

To study \source we primarily use archival \nicer observations because the large number available makes it possible to extend this study over the course of several years. We limited archival \nicer observations to those with $>5$ ks before filtering, leaving us with 51 total observations. This exposure cut is chosen to guarantee all observations (regardless of state) have above $10^6$ total counts, ensuring a comparable data quality to even the lowest flux observations from \cite{moutard23}. We then calibrate all of these observations using {\sc nicerdas} 2023-08-22\_v011a and {\sc caldb} 20221001. This is done by first using {\sc nicerl2} for geomagnetic prefiltering, then followed by {\sc nimaketime} with {\sc COR\_SAX} $> 4.0$ to manage particle overshoots and {\sc KP} $<5$ to filter out high particle background. We then construct good time intervals (GTIs) using {\sc niextract-events}. To construct spectra, backgrounds, and responses we use {\sc nicerl3-spect} with the 3C50 background model. We bin the spectra using the optimal binning scheme with a required count rate of 30 counts per bin to ensure $\chi^2$ statistics are valid \citep{kaastra16}. We construct light curves using {\sc nicerl3-lc}, which normalizes the light curves to the number of active detectors.  These light curves are generated using several X-ray bands, including the total band (0.5-6.8 keV), the super soft band (0.5-1.1 keV), the soft band (1.1-2.0 keV), the intermediate band (2.0-3.8 keV), and the hard band (3.8-6.8 keV). We compile these light curves into  hardness-intensity and color-color diagrams (Figure \ref{fig:cid}), which demonstrate that the hard color of this system is largely unchanged over the course of observations. One observation ({\sc OBSID} $=$ 1050020115) contained a type-I X-ray burst, which was filtered out using a custom good time interval (GTI). The observation date of all 51 \nicer observations are shown overlaid on a long term MAXI light curve in Figure \ref{fig:maxi}, with insets that display 20 day intervals.

We also look at 6 \nustar observations of \source. These are generated by first defining source regions in a 100$''$ circle centered on the source, and another 100$''$ circle centered off the source to define a background region using ds9. The reduction of these observations is done using {\sc nustardas} v2.1.2 and {\sc caldb} 20230816. More information on all observations can be seen in Table \ref{tab:obs} in Appendix A. We plot the the \nustar light curves using 1 second time bins, and detect no bursts in any observation, so no further filtering was done.

\section{Results and Analysis} \label{sec:results}
In this section we describe the specifics of the models used in this work to study the reflection spectrum. We begin by first modeling only the continuum components, then proceed to add a reflection component to the model.

\subsection{\nicer Continuum Modeling} \label{subsec:cont}
Before using reflection models, we first test a continuum model similar to the one used in \cite{moutard23}. The models are constructed and fit in {\sc xspec ver. 12.13.1} \citep{arnaud96}. The first component is {\sc tbabs}, a multiplicative model component used to measure absorption due to neutral hydrogen in the interstellar medium (ISM), using the {\sc wilm} cross section \citep{wilms00} and the {\sc vern} cross-section \citep{verner96}. This is followed by two multiplicative absorption edges to manage features at $\sim0.4$ and $\sim0.9$ keV, which can interfere with the \ox line. These features are potentially astrophysical in origin, owing to the ISM along the line of sight \citep{pinto13}. The inclusion or exclusion of these edges may interfere with measurements of the \ox feature. For the sake of consistency with \cite{moutard23}, we continue this analysis with two edges, but we discuss the implications further in Section \ref{subsec:edges}. The higher and lower energy edges roughly correspond to the Fe L and C K edge respectively, though the edge at lower energy has a centroid that sits near or below the band used, so it is less confidently measured. We are deferring to previous studies of this source and similar sources which use such edges \citep{madej14,ludlam20,moutard23}. After fully modeling the spectra (see Section \ref{subsec:refl} for more details), we find the edges have average energies at $\sim0.40$ and $\sim0.87$ keV, with respective optical depths of $\sim0.85$ and $\sim0.18$. The lower energy edge is roughly consistent with a known nitrogen detector edge, while the higher energy edge sits in the region of a Fe-L or Ne-K edge. The remaining model components are {\sc bbody} to account for thermal emission from the NS, {\sc diskbb} to account for thermal emission from the disk, and {\sc cutoffpl} to account for the nonthermal emission from the corona. We use {\sc cutoffpl} rather than a standard {\sc powerlaw} because many observations in the high-soft state display a high energy cutoff within the bands observed \citep{degenaar18}. Using this while allowing the cutoff energy to increase beyond the band provided by \nicer allows us to account for spectra both with and without this energy cutoff, allowing our model to be agnostic to the spectral state. The final model used to fit the continuum is {\sc tbabs*edge*edge*(bbody+diskbb+cutoffpl)}. All of the models are fitted in the $0.45-9$ keV range for consistency with \cite{moutard23}.

\begin{figure}[h!t]
\begin{center}
\includegraphics[width=0.48\textwidth, trim = 0 0 0 0, clip]{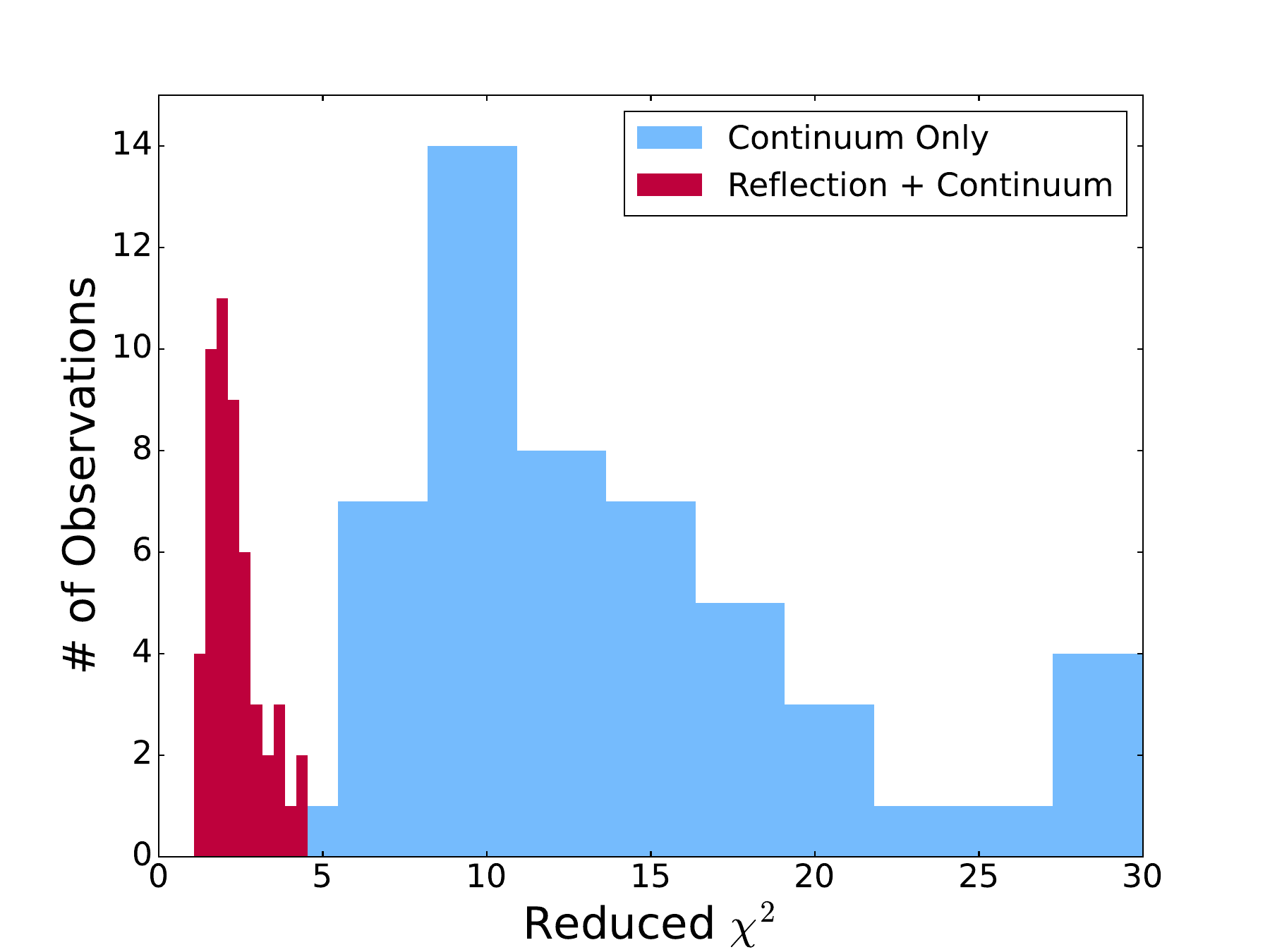}
\caption{A comparison of the reduced $\chi^2$ values for the observations when fit only with a continuum, versus when fit with a reflection component. We see that a reflection component significantly improves the reduced $\chi^2$ statistics for all observations. This should not be understood as a robust statistical comparison of the models, but rather a quick reference to determine whether a reflection component significantly improves the fit quality. The values shown here correspond to an average $\chi^2$ improvement of $\sim1566$ for the additional 6 degrees of freedom. Note: the two models are binned differently for readability. }
\label{fig:chihist}
\end{center}
\end{figure}

With the model in place, we proceed to fit all 51 \nicer spectra. We use $\chi^2$ statistics with Churazov weighting. Churazov weighting is similar to standard $\chi^2$ statistics, but takes into account adjacent bins to help prevent overfitting \citep{churazov96}. However, this can also lead to higher values of $\chi^2$ statistics. If we were using standard weighting, the reduced $\chi^2$ values reported would often be above acceptable levels, but the values found in this analysis are consistent with those from \cite{moutard23}. With the fit in place we then calculate the errors on continuum parameters by using the fitted model as a starting point for a Markov-chain Monte Carlo fit with 100 walkers, a burn-in of 100000, and a chain length of 10000. This is done using the {\sc chain} command in {\sc xspec}. To demonstrate the quality of these fits, we show a histogram of reduced $\chi^2$ values in Figure \ref{fig:chihist}, and compare the continuum-only model to a model with a reflection component added (as discussed in Section \ref{subsec:refl}). The inclusion of the reflection component leads to an improvement of in $\chi^2$ values of $\sim1566$ for 6 degrees of freedom. A table of all continuum values can be found in Appendix B. 

\subsection{\nicer Reflection modeling} \label{subsec:refl}

\begin{figure*}[t!]
\begin{center}
\includegraphics[width=0.98\textwidth, trim = 0 9 0 5, clip]{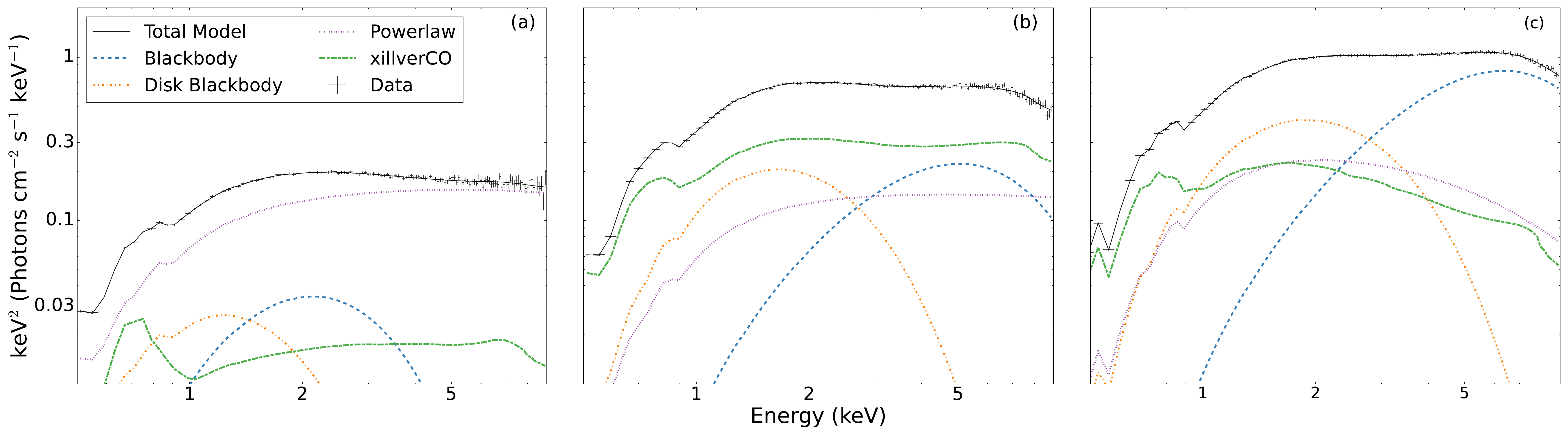}
\caption{ An example of 3 unfolded \nicer spectra with the relevant model components. We show here (a) the lowest flux observation, obsid = 4701010301, (b) an intermediate flux observation, obsid = 1050020236, and the highest flux observation, obsid = 1050020252. These display the range of fluxes we see across these observations, as well as the most relevant components for various fluxes. In (a), we see the power law is dominant, and remains relatively flat out to the highest energies, while in (c) the cutoff and preference for a stronger thermal component becomes more apparent. In each, the \ox feature is visible around 0.7 keV.}
\label{fig:specs}
\end{center}
\end{figure*}
With the continuum models in place we then proceed to add a reflection component. This reflection component is modeled using {\sc relconv*xillverCO}. As mentioned previously, \xillverco is an adaptation of \xillver with high carbon and oxygen abundances to account for UCXB features. {\sc relconv} is a relativistic convolution kernel that accounts for the broadening of the features in \xillverco. In {\sc relconv} we tie the two emissivity indices $q_1=q_2=q$. This makes the parameter $R_\mathrm{BR}$ redundant, so we fix it to 500 \rg. Since we are dealing primarily with the inner disk region, we also fix the outer radius to 990 \rg, as we do not expect this parameter to affect our results. We fix the dimensionless spin parameter $a$ to 0. The limb darkening parameter, used to control whether the model accounts for limb darkening in the source, is also set to 0. This leaves us with 3 free parameters in {\sc relconv}: $q$, as mentioned above, the inclination $i$, and the location of the inner disk radius \rin, displayed in this paper in units of \risco. For \xillverco and {\sc relconv}, we fix the inclination to $55^\circ$; this value is an average of measured values from previous works \citep{madej14,ludlam19a}, and serves to reduce the number of free parameters. The same was done in \cite{moutard23}, which finds that the parameters do not vary greatly when the inclination is allowed to be free. We tie the powerlaw index and energy cutoff in \xillverco to those from {\sc cutoffpl}. The remaining free parameters are the normalization \footnote{See \cite{dauser16} for a detailed explanation of the normalization}, the carbon-oxygen abundance $A_\mathrm{CO}$, the frac parameter, and the temperature $kT_{\mathrm{xill}}$. The frac parameter describes the ratio of the illuminating flux from the power law to that of the blackbody which emerges from the disk at the point of reprocessing. $kT_{\mathrm{xill}}$ is the temperature of that emergent blackbody. The ionization $\log\xi$ is not a parameter in \xillverco, but can be calculated using

\begin{equation}
    \xi = \frac{4\pi}{n} F_x
\end{equation}

where $F_x = {\rm frac} \times \sigma T^4$ with frac being the same as the frac parameter, $\sigma$ being the Stefan-Boltzmann constant, and T being the temperature from $kT_{xill}$ \citep{garcia13}. $n$ is the disk number density, which is fixed to $10^{17}$ cm$^{-3}$ in \xillverco.

We add the reflection component to the continuum model, and then fit using {\sc xspec}. In order to calculate errors, we then repeat the process described for the continuum in Section \ref{subsec:cont}. We also calculate the fluxes (both overall and for each component) using the multiplicative model component {\sc cflux} once the models are fit. The results of fitting all of these models can be found in Appendix B. We measure the equivalent width (EW) of the \ox feature by applying a {\sc diskline} model component to the continuum model. {\sc diskline} is a model designed to fit emission features in accretion disks \citep{fabian89}. We fit the spectrum and use the {\sc eqwidth} command in {\sc xspec} to extract the EW and 90\% confidence limit errors. We fix the inclination and outer disk radius to the same values used in \xillverco, and allow the emissivity index, inner disk radius, line energy, and normalization to be free. We use the same chain method described above to calculate errors, and the results can be found in the Table \ref{tab:refl_flux}. These equivalent widths are consistently around $\sim50$ eV, which is similar to the EW of other measured \ox lines, and is comparable to measured EWs for Fe lines in LMXBs \citep{madej11,cackett10}. The centroids of these lines are also consistently near 0.7 keV, as anticipated.

\subsection{\nustar Spectra} \label{subsec:nustar}
\begin{figure}[t!]
\begin{center}
\includegraphics[width=0.48\textwidth, trim = 30 0 0 0, clip]{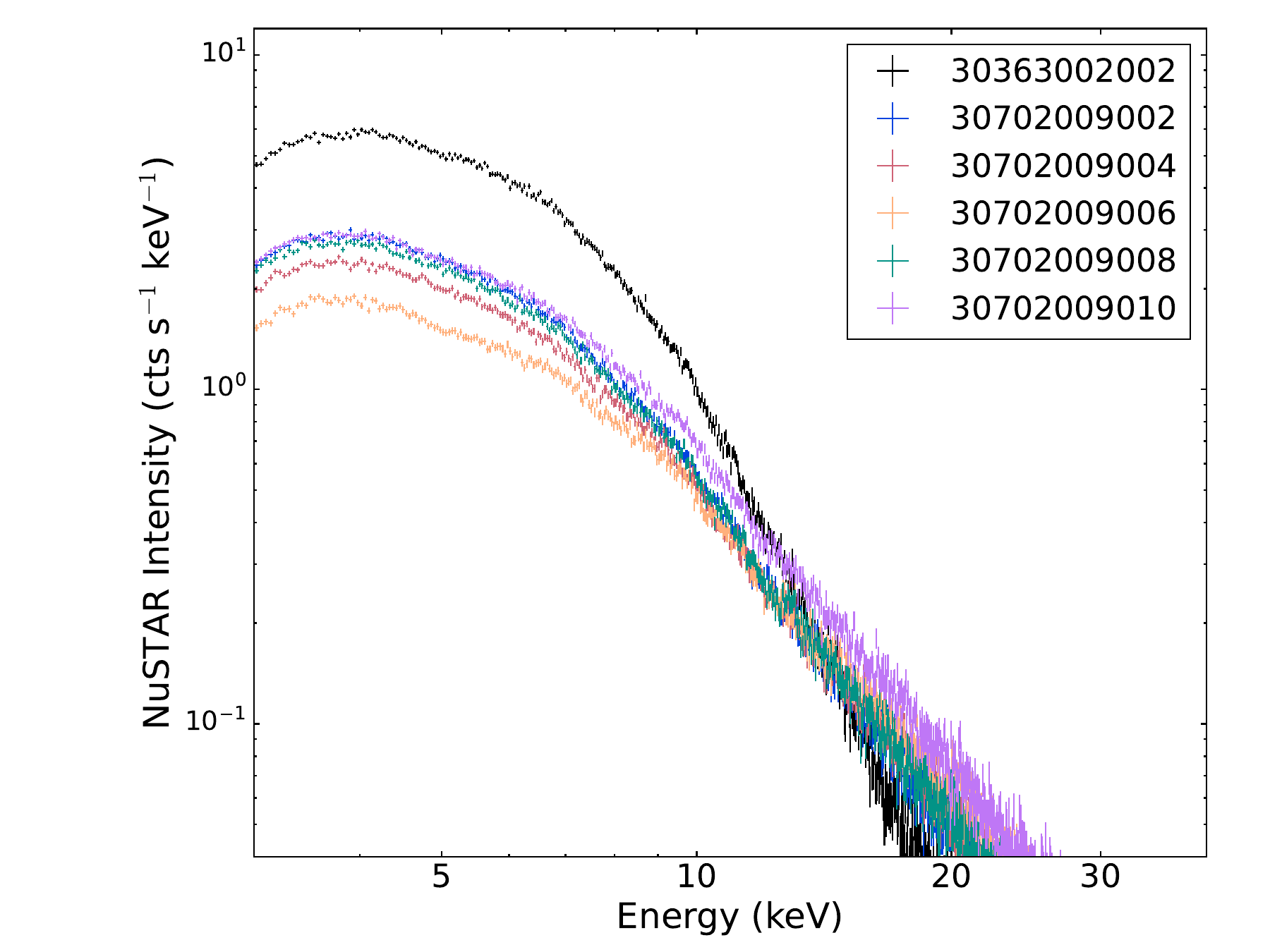}
\caption{Shown here are the 6 archival \nustar spectra. The 5 most recent spectra have comparable slopes in the higher energy bands, but the earliest observation, 30363002002, is noticeably different. This means that we can not effectively use \nustar observations which are not simultaneous with \nicer observations to constrain continuum parameters for the broader study.}
\label{fig:nustar}
\end{center}
\end{figure}

We use the 6 \nustar spectra primarily to test whether the high energy components of the continuum remain constant over time. These spectra unfortunately do not probe the softest regions of the X-ray band. This means they can not be used to measure the \ox feature, and so they are used primarily for constraining high energy continuum features. A similar power law index was seen in all observations of \cite{moutard23}, so we see if this remains true for observations separated further in time. If that were the case, we could use this fact to fix certain parameters, which would be useful in dealing with the large number of spectra involved in this study. However, we see that the slope of the hard X-ray spectrum varies over time (see Figure \ref{fig:nustar}). This means that we can not use this to constrain and fix the slope of the power law component, but does indicate that the spectral state is varying over time.

To study the constraints we can place using the \nustar spectra only, we use a similar model to the one described above and model the spectra from 3 to 30 keV, with the spectra binned optimally again with a minimum of 30 counts per bin. This band is chosen to be consistent with \cite{moutard23}. The main differences between the model used for the \nustar spectra and the \nicer spectra is that we omit the {\sc edge} components, as those model features in the low energy, and we add a constant component to account for differences between \nustar FPMA and FPMB. The constant is fixed to unity for FPMA and allowed to be free for FPMB. Since {\sc TBabs} primarily affects the lowest energy bands, we fix the value of $N_H$ to $0.4\times10^{22}$ cm$^{-2}$, an approximate average of the reflection model. We find that the results of certain parameters, such as the power law and the blackbody, are comparable to \nicer values and to those in literature \citep{moutard23}. However, other parameters vary more significantly, with predictions of extremely truncated inner disks, unrealistically low (and sometimes negative) emissivity indices, low $A_{CO}$, and occasionally poorly constrained values from {\sc diskbb}. This highlights the utility of instruments with lower energy band passes for UCXB studies, and demonstrates the ability of \nustar to help constrain continuum parameters. The results of these fits can be found in table \ref{tab:nustar_refl} in Appendix C.

\section{Discussion}\label{sec:disc}
\subsection{Flux and Inner Disk Radius}\label{subsec:flux}
\begin{figure}[t!]
\begin{center}
\includegraphics[width=0.48\textwidth, trim = 20 0 20 20, clip]{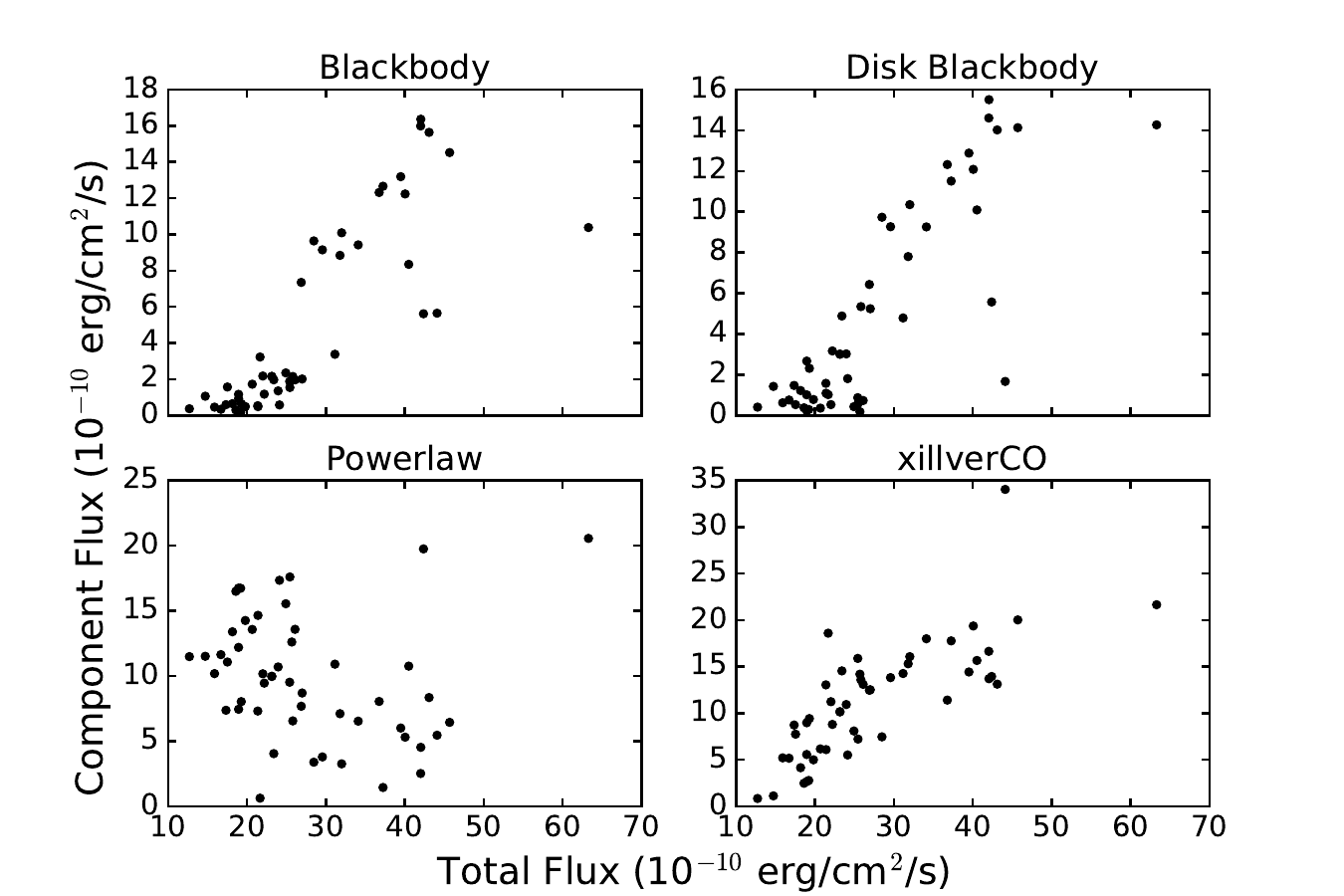}
\caption{Shown are the fluxes of each component for each observation plotted against the overall flux of the system. In general, the component fluxes are well-correlated with the overall flux (spearman rank $\geq 0.75$), with the exception of the powerlaw flux, which is anti-correlated (spearman rank $\sim -0.5$). }
\label{fig:fluxes}
\end{center}
\end{figure}

\cite{moutard23} discusses \source in terms of the overall flux behavior. In that analysis, the flux of all model components scale with the overall flux except the flux of the powerlaw component, which seems to be anti-correlated. The decrease in flux of the \xillverco component as the illuminating powerlaw increases is explained by a truncated disk during the lowest flux observation. This can also be explained by the fact that the fraction of emitted coronal photons to those reflected 
 decreases as the height of the corona increases. If the coronal height decreases as the source gets softer and the powerlaw becomes weaker, this means more photons illuminate the disk and are reflected, resulting in a greater contribution from \xillverco \citep{dauser16}. We find in this analysis that the flux patterns persist over longer time scales as well. We show in Figure \ref{fig:fluxes} that, in general, the flux of each component is increasing with overall flux except the powerlaw component. This is consistent with the well-documented low-hard and high-soft states, where higher flux observations are more thermally dominated and lower flux states are more dominated by non-thermal coronal emission. These correlations are supported by a spearman rank $\geq 0.75$ for all components except the anticorrelated power law, which is at a spearman rank of $\sim -0.5$. We also note that the anticorrelation of the powerlaw flux and \xillverco flux discussed above is supported by a spearman rank of $-0.6$.

\begin{figure}[t!]
\begin{center}
\includegraphics[width=0.48\textwidth, trim = 20 0 20 20, clip]{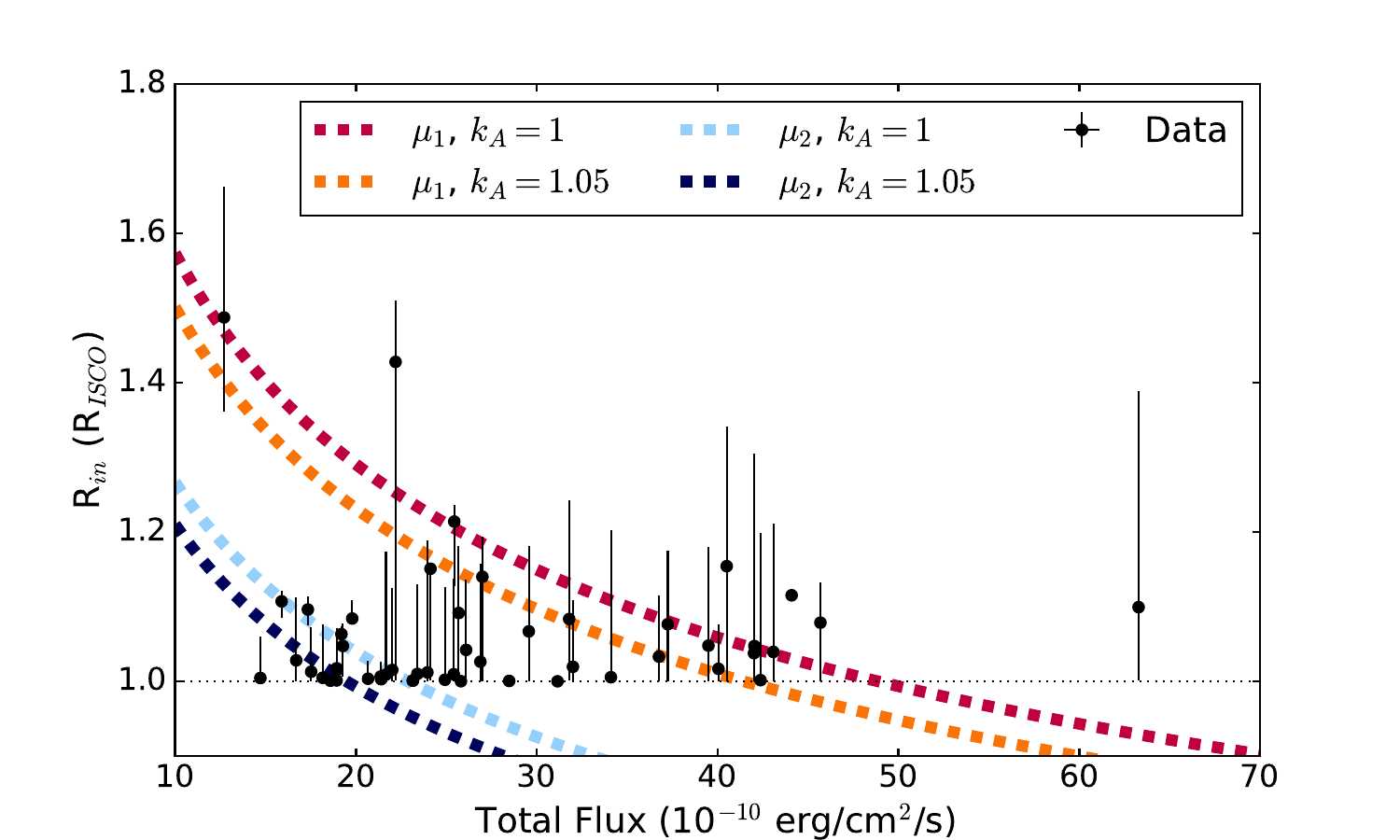}
\caption{The overall flux of the system is poorly correlated with the inner disk radius. We see the lowest flux states display some degree of truncation as stated in \cite{moutard23}, but there are also higher flux states which are truncated just as heavily, if not moreso. To test whether the observed disk truncation is caused by magnetic fields surrounding the NS, we plot the magnetosphere radius for various limits. Assuming our magnetic field upper limit (red dashed line) is well characterized by the most lowest flux point does not yield a magnetosphere radius consistent with most measured \rin values. If instead we use a different low flux, seemingly truncated observation to characterize the magnetic field strength (blue dashed line), we find a slightly better relation, though it still can not be concluded that the magnetic field is the sole source of inner disk truncation in this system. $k_A$, $f_{ang}$, and $\eta$ are all degenerate, so it is possible to make these magnetic field radii very small or very large, even using reasonable values for this parameter.}
\label{fig:rins}
\end{center}
\end{figure}

The truncated inner disk at the low-flux state is pointed to as an analogy for BH-LMXB systems, which tend to show a larger value of \rin for low luminosities (which is a correlated with the mass accretion rate, \citealt{done07,tomsick09,garcia15}). However, over longer timescales, \source does not appear to strongly show any correlation between inner disk radius and flux, aside for a slight preference for truncation in the lowest  flux states. This relationship is shown in Figure \ref{fig:rins}, and even though the lowest flux states display some disk truncation, we see a random assortment of higher flux observations with equally or more truncated disks. We test a handful of lower flux observations to see if the disk truncation can be accounted for by the magnetic field. In some NS LMXBs, the magnetic field is able to truncate the disk at the Alfv\'{e}n radius ($r_A$), where the energy density of the magnetic field equals the energy density of the accreting material. To do this we calculate the strength of the equatorial magnetic field during our most truncated observation using the following equation from \cite{ludlam20}:

\begin{multline}
    B = 3.5 \times 10^5\ k_A ^{-7/4} x^{7/4}\left( \frac{M}{1.4\ M_{\odot}} \right)^2 \left(\frac{10\ \rm km}{R_{\rm NS}}\right)^3 \\
    \times \left(\frac{f_{ang}}{\eta} \frac{F_{bol}}{10^{-9}\ {\rm erg s^{-1}cm^{-2}}} \right)^{1/2} \frac{D}{3.5\ \rm kpc}\ \rm G
\end{multline}

where the efficiency $\eta$ is assumed to be 0.2 \citep{sibgatullin00}, the conversion factor $k_A$ and the angular anisotropy $f_{ang}$ are set to unity \citep{ludlam19a}, and the distance $D$ to \source is 3.3 kpc (determined using GAIA data in \citealt{arnason21}). We use canonical NS values $M = 1.4 M_\odot$ and $R_{NS} = 10$km, and the $0.5-50$ keV flux and the inner disk radius (in $R_g$) are used for $F_{bol}$ and $x$ respectively. If we use the lowest flux observation to determine our magnetic field strength, this provides an upper limit on the magnetic field of $B_1 \leq 0.4\times10^8$G. We convert this value to a magnetic dipole moment (again assuming the values above) and find $\mu_1 = 3.8\times10^{25}$ G cm$^{-3}$. We use this value and mass accretion rate $\dot{m} =  L/\eta c^2$ where $L$ is the luminosity during that observation to calculate  

\begin{equation}
    r_A = (\frac{\mu^4}{GM\dot{m}^2})^{1/7}
\end{equation}

where G is the gravitational constant. This radius is plotted as a red dashed line on Figure \ref{fig:rins}. We see that this radius is generally too high to explain the inner disk radius in this case. These radii are generally too large to assume that the disk is truncating due the magnetic field. If instead we use the {\it next} lowest flux observation with some indication of truncation to determine our magnetic field strength, we find $B_2 \leq 0.3\times10^8$G and $\mu_2 = 2.6\times10^{25}$ G cm$^{-3}$, for which we plot $r_A$ as a blue dashed line in Figure \ref{fig:rins}.

If instead we assume $k_A = 0.5$ \citep{long05},  $\mu$ increases by roughly a factor of three, which can drastically increase $r_A$, especially in the case of $\mu_1$. This would extend $r_A$ beyond the truncated inner disk, meaning truncation can not be caused by the magnetic field. If instead we assume $k_A = 1.05$ (below the theoretical upper limit of 1.1, discussed by \citealt{ibragimov09}), we find that in some cases, the magnetosphere is consistently at or below the disk truncation. In such a case, it is possible that disk truncation is caused by the magnetic field during the low flux states. We can see similar changes to $r_A$ by assuming different values of $f_{ang}$, (though it is expected to be near or above unity \citealt{ibragimov09}) or $\eta$ (which can vary depending on the rotational frequency of the NS). Because of this, we can not confidently say that the low flux disk truncation is caused exclusively by the magnetic field. 

Another possibility for the cause of the disk truncation is disk depletion at low mass accretion rates. However, this was explored for \source in \citep{moutard23}, which determined that the timescale required does not correspond with the cadence of observations. It should be noted that the degree of truncation is quite small, only extending up to $\sim1.5$ \risco (compare this to truncation up to $\sim10$\risco as seen in Figure 5 of \citealt{ludlam24}). This small degree of truncation, paired with the limitations of \xillverco at high flux (discussed in more detail in Section \ref{subsec:xillvercolimit}) makes drawing conclusions about the cause of the truncation difficult.
\subsection{Emissivity Index}\label{subsec:emissivity}
The emissivity index $q$ is a value that represents the power law index defining the illumination of the disk as a function of radius, $r^{-q}$. In the classical limit, $q=3$, but this can be affected by multiple factors, such as the geometry of the illuminating corona and relativistic light bending \citep{wilkins18}. One can use the emissivity index as a stand in for how heavily light is curved toward the disk; in BH systems, this index is often much higher as the gravitational effects curve more of the emission to the regions of the disk closest to the BH. This leads to a steeper fall-off in illumination near the inner disk ($q \gtrsim 6$ in the case of rapidly spinning BHs), before returning to the classical limit at the outer disk \citep{wilkins12}. 

\begin{figure}[t!]
\begin{center}
\includegraphics[width=0.48\textwidth, trim = 140 190 70 0, clip]{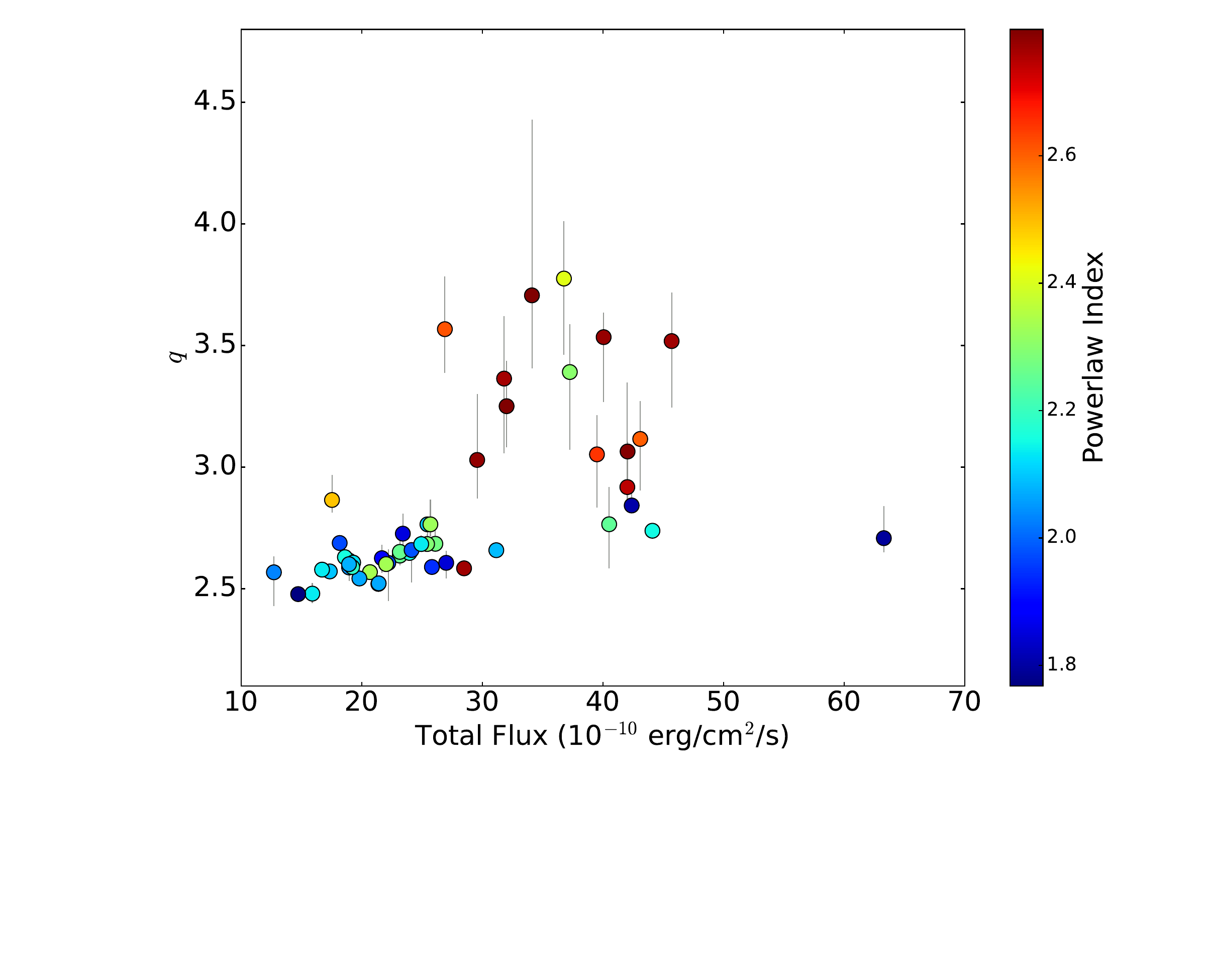}
\caption{Shown is the emissivity index $q$ plotted against the total unabsorbed flux of the system. Flatter emissivity profiles are generally consistent with the system in a low-hard state, while steeper values of $q$ correspond with the high/soft state. This is potentially explained by a corona which is extending spatially during low flux states.}
\label{fig:q}
\end{center}
\end{figure}

For NS systems, these values are typically closer to the classical limit. In this study we use only one emissivity index to minimize the number of free parameters, shown in Figure \ref{fig:q}. By using only one value for $q$, we are essentially adopting an average value of the inner and outer emissivity indices. Values of $q < 3$, or a flattening of the emissivity index, may be indicative of a spatial extension of the illuminating corona. This may mean a vertical extension above the axis of the disk or a radial extension as the corona covers the inner region of the disk. We see in Figure \ref{fig:q} that for low flux sources, $q$ is generally consistent with a slightly flatter profile, while in higher flux states, the index tends to increase. This is also correlated with the spectral state of the system, as shown by the color bar representing the index of the power law. A larger power law index is indicative of a softer spectrum, which is expected for higher flux states. 

This relationship is potentially analogous to what is seen by \cite{kara19} for the BH transient system MAXI J1820+070. In that study, they propose a geometry wherein the corona extends vertically above the disk as the spectrum becomes harder. Similarly, \cite{buisson19} finds a similar result for the same source, indicating that the height of the illuminating corona decreases as the spectrum softens. While NS and BH XRBs are not exactly the same, the accretion processes are similar, so an analogy between these systems is not unreasonable. \source is known to contain a relativistic jet during its hard states, though the jet alone is likely not a viable mechanism to produce hard X-rays, and would also require some other source of Comptonized photons \citep{migliari10}. It is unlikely the jet is turning on and off on the time scale of weeks, following the quasi-periodic flux in the MAXI light curve. This provides evidence that the jet is not the driving source of the coronal photons, though it may contribute in some cases.

Another possible geometry of the spatially extended corona is one that expands radially rather than vertically above the disk. If at some point the disk becomes highly magnetized, this could provide the fields needed to produce the Comptonized corona which would then illuminate the disk from directly above it. This too would lead to a flattening of the emissivity index over the region where the corona sits \citep{wilkins18}. However, at least in AGN systems, this coronal geometry is predicted primarily for high-flux states \citep{wilkins14}. In \source, the flattening of $q$ appears to occur in the lower luminosity states, opposite to what is seen for systems where this geometry is well studied. Because of this, as well as the known existence of a jet for this source, we deem that geometry to be less likely. Therefore, the most likely scenario is a vertically extended corona. However, as mentioned in Section \ref{subsec:flux}, the limitations of \xillverco make drawing broad conclusions about the transition into high-soft states difficult.

\subsection{Limitations of \xillverco}\label{subsec:xillvercolimit}
It is worth noting that \xillverco assumes disk illumination by a non thermal Comptonized component. This works well for UCXBs in the low-hard state, but may begin to falter when faced with disks that are illuminated by thermal emission from near the NS itself. This regime is represented by the the high-soft state, where we see $q$ increase. This may serve to explain some of the discrepancies in the values of $q$ for the few high flux but apparently soft points in Figure \ref{fig:q}. These observations don't seem to agree with the overall trend of $q$ increasing in the high flux states. These outlier observations are also among those with the worst reduced $\chi^2$ value. The self consistent reflection model \relxill also assumes illumination by a Comptonized corona, but the extension of the model, \relxillns, assumes thermal illumination. Similarly, a version of \xillverco with thermal illumination would be useful to describe the source in all states. 

\begin{figure}[t!]
\begin{center}
\includegraphics[width=0.48\textwidth, trim = 140 190 70 0, clip]{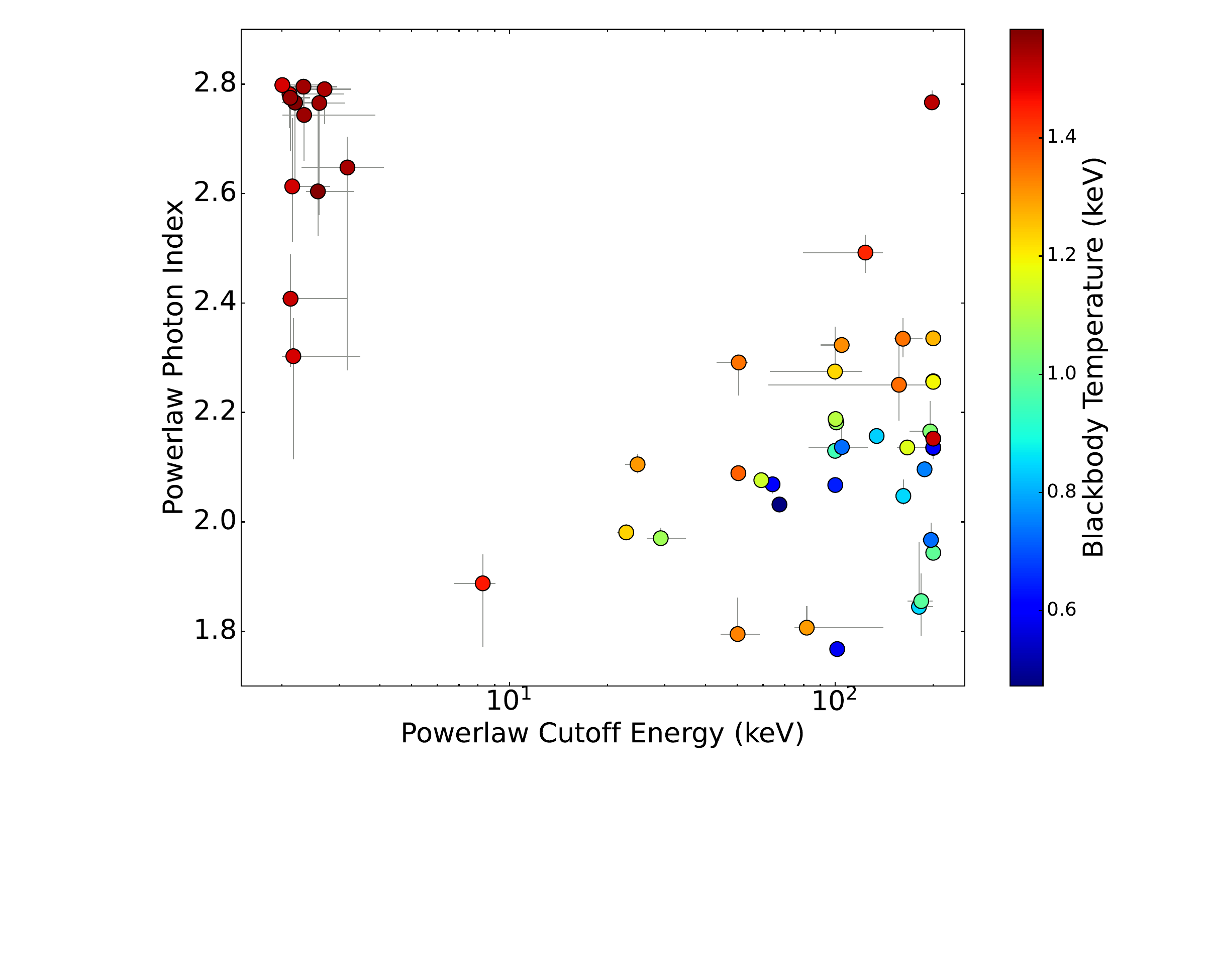}
\caption{Shown here is the powerlaw photon index plotted against each spectrum's powerlaw cutoff energy, color coded by the blackbody temperature. While many observations demonstrate the expected behavior of a soft spectrum with a high blackbody temperature, some of the highest temperature blackbodies are coincident with steep illuminating powerlaws with very low energy cutoffs. These observations are at the edge of the limits of \xillverco, which assumes illumination by a powerlaw, even if the disk is being physically illuminated by the thermal NS boundary layer. In these cases, the model attempts to force the illuminating powerlaw into a shape resembling that of a blackbody.}
\label{fig:limit}
\end{center}
\end{figure}

Figure \ref{fig:limit} demonstrates one such limit of \xillverco. The bottom right region of the plot shows the coolest blackbody temperatures being associated with shallow photon indices and high energy cutoffs, as expected for harder spectral states. As the blackbody temperature increases, we see either an increase in the power law index or a decrease in the cutoff energy (the two parameters are somewhat degenerate, especially when the energy cutoff sits beyond the \nicer band used in the study). However, in the top left region of the plot, we see that the observations with the highest blackbody temperatures demonstrate a very soft powerlaw spectrum. These represent a regime where the illuminating power law shifts to the lowest energies and cuts off quite steeply. A steep, absorbed power law with a very low energy cutoff mimics the shape of a blackbody. Figure 1 in \cite{garcia22} demonstrates the difference between \xillver and \xillverns. We see therein that even when the incident spectrum mimics the shape of a blackbody illuminating the disk, the resulting reflection spectrum depends heavily on the input model used. For this study, that means that using a model with the incorrect input spectrum can cause some of the modeled parameters to tend towards unphysical values.

\subsection{Absorption Edges}\label{subsec:edges}
\begin{figure}[t!]
\begin{center}
\includegraphics[width=0.48\textwidth, trim = 0 0 0 0, clip]{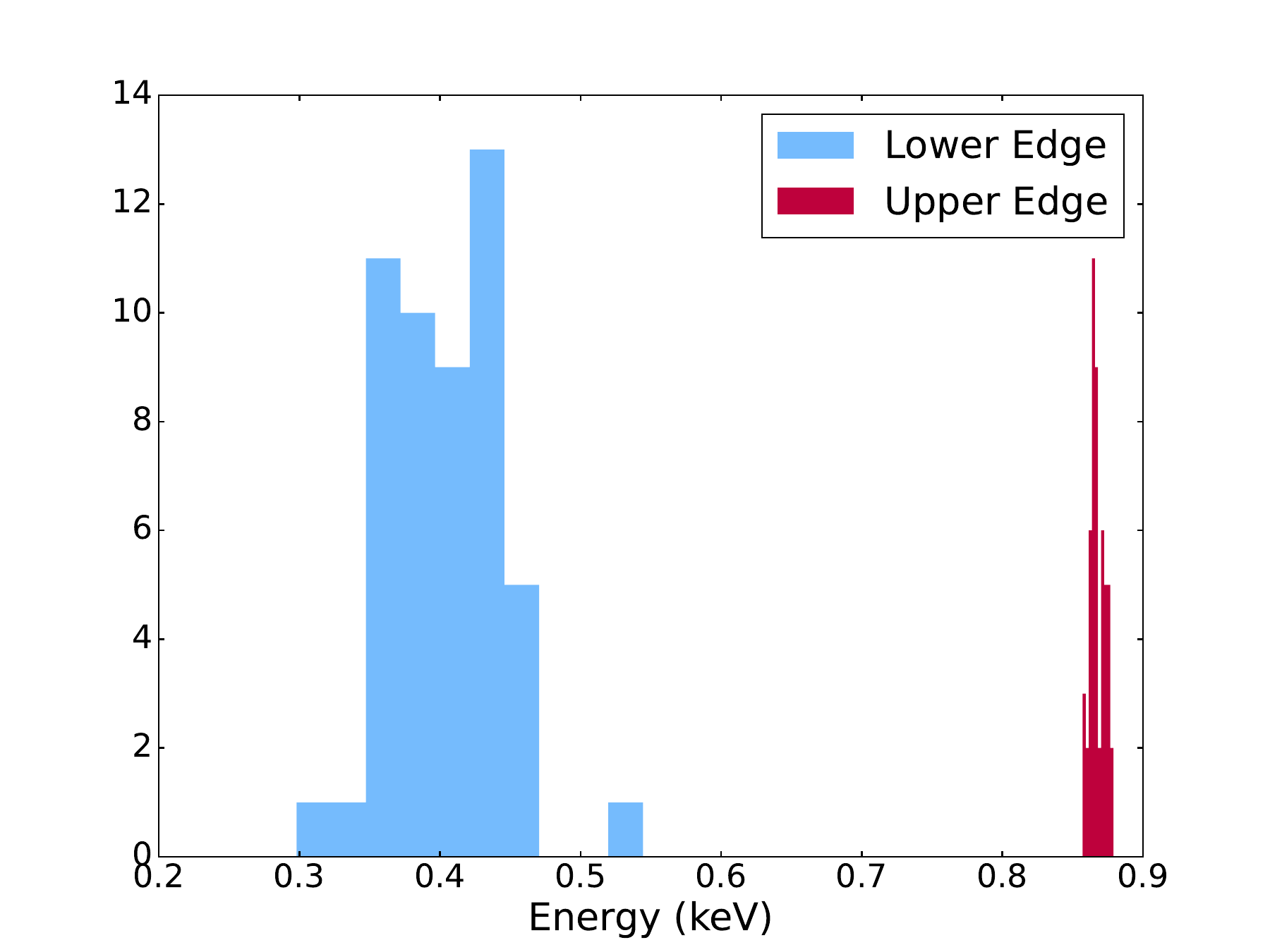}
\caption{The distribution of edge energies for the model containing both edges. We see that the higher energy edge near 0.87 keV (consistent with a Ne-K or Fe-L edge) is very well constrained, while the lower energy edge around 0.4 keV (consistent with a N detector edge) has a significant amount of spread. This is to be expected for a feature that sits at or below the given \nicer band, but should be addressed more carefully in future studies with higher resolutions and improved collecting area at the softest energies.}
\label{fig:edges}
\end{center}
\end{figure}

In the model we include two absorption edges in the lowest energies to account for apparent features at roughly 0.4 and 0.87 keV (corresponding roughly to a N detector edge and a Fe-L or Ne-K edge, respectively.) The inclusion of absorption edges surrounding the \ox feature at $\sim0.7$ keV can cause some interference with the measurements of the feature itself. Between the two edges, the higher energy edge is much more well constrained, as shown in Figure \ref{fig:edges}. The lower edge is less constrained, but that is to be expected as it sits at or below the lower end of the \nicer band in use. As a result we initially test fits using only one edge centered near 0.87 keV, while keeping the rest of the model the same. The reduced $\chi^2$ of the fits worsen marginally when the lower energy edge is removed. We find that the distributions of most parameters do not vary greatly, with some exceptions. $A_{CO}$ tends to increase dramatically when one edge is removed, offering values up to 20 times higher than those from models with two edges. These carbon-oxygen abundances, reaching up to several hundred times solar is inconsistent with \cite{madej14} when accounting for the fact that the abundances in that paper are artificially increased in that version of the model. \cite{moutard23} finds the value of $A_{CO}$ to be near 25, which is broadly consistent with \cite{madej14} (though it should be noted that 2 edges are also used in \citealt{moutard23}).

We also find that the model with only one edge tends to fit the soft state using a harder powerlaw photon index than is physically reasonable for these states, although with a low energy cutoff. This model trends towards $\Gamma \leq 1.2$ during the softest states, which is only observed in the extreme hard states for NS systems \citep{ludlam16,parikh17}.  While the trends between the models do not match, the conclusions drawn from Figure \ref{fig:limit} remain: in the soft state, the powerlaw component cutoff energy shifts to imitate a thermal component. With only one edge, the trend seen in Figures \ref{fig:q} and \ref{fig:rins} is also less consistent, with more spread in these parameters. We further test the validity of the edges by fixing their energies at 0.4016 keV and 0.8669 keV, corresponding to the laboratory values of the N edge and the Ne-K edge. We choose the Ne-K edge value over the Fe-L edge as it is much closer to the average free value of 0.8677. Similarly to the test with one edge, while some outliers do appear, the overall distributions of parameters remain similar to the free energy fits. Again we see some increase in  $A_{CO}$, though not as dramatically as in our single edge test. We also see some discrepancies in the disk blackbody temperature, which is unsurprising as the central energy often coincides with the lower energy edge. However, other key parameters studied over the course of this paper, such as \rin and the emissivity index, maintain very similar values after freezing the edge energies. For completeness, we also test a handful of observations with no edges at all. When not accounting for the well constrained higher energy edge, we found that the reduced $\chi^2$ increases drastically for every observation checked, sometimes by a factor of greater than 5. Because of this, we do not pursue such a model further. Without a careful study of these absorption edges at low energy, it is difficult to say which model best explains the physical nature of the system. As a result we primarily focus on the model that is a more direct comparison with \cite{moutard23}, but this does encourage a low energy, high resolution study of this source to better constrain the edge components.

\section{Conclusion}\label{sec:conc}
In this study, we examine the long term behavior of \source via 51 archival \nicer spectra. These spectra cover a range of time spanning over 5 years, and capture the source in various different spectral and flux states. After using \xillverco, a reflection model designed for UCXBs, we are able to say the following about this source:

\begin{itemize}
    \item The flux behavior is consistent with \cite{moutard23}, which follows along the conventional high-soft and low-hard states. That is, as the overall unabsorbed flux of the system increases, so do the fluxes of most of the individual model components. The exception is the flux of the powerlaw, which represents the comptonized corona. This component decreases as the source gets brighter. 
    \item The relationship between the flux and \rin is seemingly uncorrelated, but displays some agreement with \cite{moutard23}, which finds that the lowest flux observations appear to show a truncated disk. This is true in some cases, but drawing any conclusion about the behavior of the inner disk is difficult with the given data. 
    \item The emissivity index tends to increase during the high-soft state when fit with \xillverco. This is indicative of the coronal extent increasing during harder states. A jet may exist during these states, and could provide some contribution to the illumination of the disk from above the NS itself, but it is not likely the driving source. In softer states, the illuminating component may recede to be closer to the NS, causing the light to be curved more steeply toward the inner region of the disk. The picture of a corona extending further from the compact object as the spectrum hardens is also analagous to certain BH XRB systems.
    \item The reflection model used here, \xillverco, may begin to falter in high flux states, where disk illumination from a nonthermal corona gives way to thermal illumination from a NS boundary layer. In these cases, the model appears to shift the incident powerlaw to the absolute lowest energies, mimicking the shape of a thermal illuminating component. During such states, the reflection model depends heavily on the correct illuminating component, and therefore makes drawing conclusions about the source in the high flux state difficult. This combined with a poor characterization of the lowest energy absorption edges leads to a model that is not complete when discussing UCXBs.
\end{itemize}

This long term study of \source paints a picture of a jetted UCXB which behaves similarly to other NS or BH LMXBs. The hard component appears to decrease in flux as the source gets brighter, and the inner disk truncation is poorly correlated with luminosity. The emissivity index provides a clue that a jet may contribute to the corona in hard states, while the illumination geometry shifts to be closer to the NS in softer states. Future high resolution studies of reflection features with missions such as XRISM could help to better constrain the emissivity index and therefore the geometry of the illuminating component. This, paired with simultaneous radio measurements of the jet could also be enlightening to determine the extent of the jet's contribution to the corona as the spectrum changes between states. Polarimetry instruments such as IXPE can be used to probe coronal geometry in some systems, and could potentially help to constrain our predictions of the extent of the corona in the soft states \citep{krawczynski22}. Also useful for this analysis would be a UCXB model which accounts for disk illumination by the NS itself rather than a comptonized corona. 
\\
\\
{\it Acknowledgements:} 
This work is supported by NASA under grant No. 80NSSC22K0054. This research has made use of MAXI data provided by RIKEN, JAXA and the MAXI team \citep{matsuoka09}. This research has made use of data and/or software provided by the High Energy Astrophysics Science Archive Research Center (HEASARC), which is a service of the Astrophysics Science Division at NASA/GSFC. This research has made use of the \nustar Data Analysis Software (NuSTARDAS) jointly developed by the ASI Science Data Center (ASDC, Italy) and the California Institute of Technology (USA).

\bibliographystyle{aasjournal}
\bibliography{bibliography}
\appendix
\setcounter{table}{0}
\renewcommand{\thetable}{A.\arabic{table}}

\section{Observation Information}
This section of the appendix summarizes all observations from \nicer and \nustar used in this analysis.
\begin{table}[h!]

\caption{\source Observation Information}
\label{tab:obs}

\begin{tabular}{lll|lll}

Observation ID & Obs.\ Start (UTC) & Exp. (ks) &Observation ID & Obs.\ Start (UTC) & Exp. (ks) \\
\hline
\multicolumn{6}{c}{\nicer}\\
\hline
1050020115 & 2017-09-07 00:52:18 & $9.7$  & 1050020225 & 2018-10-16 00:37:00 & $14.8$\\ 
1050020116 & 2017-09-08 01:24:00 & $7.4$ & 1050020226 & 2018-10-17 00:04:00 & $12.0$\\ 
1050020117 & 2017-09-09 05:25:30 & $7.2$ & 1050020229 & 2018-10-20 02:03:00 & $5.1$\\ 
1050020118 & 2017-09-10 03:01:02 & $6.0$  & 1050020230 & 2018-10-20 23:42:20 & $11.6$\\ 
1050020119 & 2017-09-11 00:38:30 & $9.7$ & 1050020232 & 2018-10-22 23:46:20 & $11.5$\\ 
1050020121 & 2017-10-19 01:20:40 & $4.8$ &  1050020236 & 2018-10-31 02:21:21 & $4.4$ \\
\end{tabular}

\end{table}

\setcounter{table}{0}
\renewcommand{\thetable}{B.\arabic{table}}
\section{\nicer Fit Parameters}
This section of the appendix summarizes all of the final fit parameters used in both the initial continuum model as well as the reflection model for \nicer.
\begin{turnpage}
    
\begin{table}[h!]

\caption{\source Results of Continuum Modeling}
\label{tab:cont}
\tiny
\centering
\begin{tabular}{l|l|ll|ll|ll|ll|lll|l}
& \multicolumn{1}{c|}{\sc tbabs} &\multicolumn{2}{c|}{\sc edge 1} &\multicolumn{2}{c|}{\sc edge 2} &\multicolumn{2}{c|}{\sc bbody} &\multicolumn{2}{c|}{\sc diskbb}  &\multicolumn{3}{c}{\sc cutoffpl}  \\
OBSID & $N_{\rm H}$ (10$^{22}$ cm$^{-2}$)&E (keV) & $\tau_{max}$ & E (keV) & $\tau_{max}$ &kT (keV) & k$_{\mathrm{BB}}$ ($10^{-3}$) & kT (keV) & k$_{\mathrm{DBB}}$ & $\Gamma$ & E$_{\mathrm{cutoff}}$ (keV) & k$_{\mathrm{PL}}$&$\chi^2$ (dof) \\

\hline \hline
1050020115 & $0.46_{-0.02}^{+0.01}$ &$0.434_{-0.002}^{+0.001}$ &$2.6 \pm 0.1$ &$0.847_{-0.001}^{+0.003}$ &$0.296_{-0.004}^{+0.01}$ &$1.44_{-0.005}^{+0.013}$ &$16.02_{-0.07}^{+0.03}$ &$0.571_{-0.003}^{+0.007}$ &$1023.5_{-62.5}^{+21.4}$ &$3.99_{-0.04}^{+0.01}$ &$0.48_{-0.01}^{+0.04}$ &$3.2_{-0.5}^{+0.1}$ &1333(143)\\ 
1050020116 & $0.12_{-0.01}^{+0.02}$ &$0.451 \pm 0.001$ &$2.7 \pm 0.1$ &$0.858_{-0.001}^{+0.002}$ &$0.399_{-0.004}^{+0.007}$ &$1.54_{-0.01}^{+0.03}$ &$12.6_{-0.8}^{+1.1}$ &$0.62 \pm 0.01$ &$432.8_{-33.4}^{+16.2}$ &$2.2 \pm 0.1$ &$483.1_{-108.7}^{+16.9}$ &$0.28_{-0.02}^{+0.03}$ &1308(131)\\ 
1050020117 & $0.17_{-0.02}^{+0.03}$ &$0.447_{-0.001}^{+0.002}$ &$2.9_{-0.2}^{+0.1}$ &$0.859_{-0.001}^{+0.002}$ &$0.43 \pm 0.01$ &$1.44 \pm 0.02$ &$6.0_{-0.7}^{+0.2}$ &$0.59_{-0.02}^{+0.01}$ &$254.3_{-27.5}^{+23.4}$ &$2.8_{-0.2}^{+0.1}$ &$455.0_{-98.0}^{+44.9}$ &$0.36_{-0.01}^{+0.03}$ &1142(127)\\ 
1050020118 & $0.15_{-0.04}^{+0.02}$ &$0.448 \pm 0.001$ &$2.9 \pm 0.2$ &$0.857 \pm 0.001$ &$0.43 \pm 0.01$ &$1.46_{-0.05}^{+0.03}$ &$4.0_{-0.7}^{+0.3}$ &$0.57_{-0.05}^{+0.02}$ &$160.6_{-21.8}^{+44.3}$ &$2.6 \pm 0.1$ &$459.5_{-230.0}^{+40.4}$ &$0.38 \pm 0.02$ &1021(124)\\ 
1050020119 & $0.27_{-0.01}^{+0.02}$ &$0.445 \pm 0.001$ &$3.0_{-0.2}^{+0.1}$ &$0.858_{-0.001}^{+0.002}$ &$0.379_{-0.006}^{+0.004}$ &$1.43 \pm 0.01$ &$11.54_{-0.06}^{+0.05}$ &$0.592_{-0.003}^{+0.006}$ &$605.0_{-36.2}^{+10.4}$ &$3.996_{-0.032}^{+0.004}$ &$1.8_{-0.1}^{+0.4}$ &$0.5_{-0.04}^{+0.03}$ &1574(132)\\
\end{tabular}

\end{table}

\end{turnpage}

\begin{turnpage}
    
\begin{table}[h!]

\caption{\source Results of Reflection Modeling: Continuum Parameters}
\label{tab:refl_cont}
\tiny
\centering
\begin{tabular}{l|l|ll|ll|ll|ll|lll}
& \multicolumn{1}{c|}{\sc tbabs} &\multicolumn{2}{c|}{\sc edge 1} &\multicolumn{2}{c|}{\sc edge 2} &\multicolumn{2}{c|}{\sc bbody} &\multicolumn{2}{c|}{\sc diskbb}  &\multicolumn{3}{c}{\sc cutoffpl}  \\
OBSID & $N_{\rm H}$ (10$^{22}$ cm$^{-2}$)&E (keV) & $\tau_{max}$ & E (keV) & $\tau_{max}$ &kT (keV) & k$_{\mathrm{BB}}$ ($10^{-3}$) & kT (keV) & k$_{\mathrm{DBB}}$ & $\Gamma$ & E$_{\mathrm{cutoff}}$ (keV) & k$_{\mathrm{PL}}$ \\

\hline \hline
1050020115 & $0.48_{-0.02}^{+0.01}$ &$0.444_{-0.003}^{+0.002}$ &$1.3 \pm 0.1$ &$0.87 \pm 0.01$ &$0.12_{-0.01}^{+0.02}$ &$1.512_{-0.004}^{+0.019}$ &$15.0_{-0.5}^{+0.1}$ &$0.642_{-0.01}^{+0.004}$ &$435.4_{-34.8}^{+16.2}$ &$2.4 \pm 0.1$ &$2.13_{-0.13}^{+1.04}$ &$0.44_{-0.11}^{+0.03}$ \\ 
1050020116 & $0.307_{-0.004}^{+0.005}$ &$0.457_{-0.001}^{+0.005}$ &$0.736_{-0.084}^{+0.004}$ &$0.879_{-0.0005}^{+0.0041}$ &$0.214_{-0.003}^{+0.001}$ &$1.3_{-0.01}^{+0.02}$ &$6.7_{-0.3}^{+0.1}$ &$0.531_{-0.036}^{+0.002}$ &$487.2_{-79.9}^{+64.6}$ &$1.81_{-0.01}^{+0.04}$ &$81.8_{-6.9}^{+58.8}$ &$0.22_{-0.05}^{+0.01}$ \\ 
1050020117 & $0.37_{-0.005}^{+0.004}$ &$0.37 \pm 0.01$ &$1.19_{-0.05}^{+0.02}$ &$0.868 \pm 0.003$ &$0.217 \pm 0.003$ &$1.2 \pm 0.02$ &$2.57 \pm 0.02$ &$0.415_{-0.002}^{+0.004}$ &$696.7_{-25.5}^{+15.2}$ &$2.257_{-0.008}^{+0.004}$ &$200.0^\dagger$ &$0.2_{-0.002}^{+0.001}$ \\ 
1050020118 & $0.367_{-0.001}^{+0.003}$ &$0.364_{-0.008}^{+0.004}$ &$0.628_{-0.005}^{+0.014}$ &$0.865_{-0.004}^{+0.003}$ &$0.194_{-0.002}^{+0.004}$ &$1.27 \pm 0.01$ &$2.06_{-0.02}^{+0.03}$ &$0.2004_{-0.0004}^{+0.0012}$ &$3006.8_{-82.3}^{+49.9}$ &$2.34 \pm 0.01$ &$200.0^\dagger$ &$0.296_{-0.003}^{+0.004}$ \\ 
1050020119 & $0.48_{-0.04}^{+0.01}$ &$0.42_{-0.01}^{+0.03}$ &$1.4_{-0.2}^{+0.1}$ &$0.872_{-0.002}^{+0.004}$ &$0.18 \pm 0.01$ &$1.5 \pm 0.01$ &$10.9_{-0.2}^{+0.1}$ &$0.693_{-0.022}^{+0.005}$ &$228.6_{-17.0}^{+32.1}$ &$2.78_{-0.06}^{+0.02}$ &$2.1_{-0.1}^{+1.0}$ &$0.2_{-0.04}^{+0.07}$ \\
\end{tabular}

\end{table}

\end{turnpage}

\begin{turnpage}
\begin{table}[h!]
\caption{\source Results of Reflection Modeling: Reflection Parameters}
\label{tab:refl_xill}
\tiny
\centering
\begin{tabular}{l|ll|lllll|l}
& \multicolumn{2}{c|}{\sc relconv} &\multicolumn{5}{c}{\xillverco} \\
OBSID & q &\rin (\risco) & A$_{CO}$ & frac & kT$_{\mathrm{xill}}$ ( 10$^{-2}$ keV) & k$_{\mathrm{\xillverco}}$ (10$^{-9}$) & $\log\xi$ &$\chi^2$ (dof) \\

\hline \hline

1050020115 & $3.8_{-0.3}^{+0.2}$ &$1.03_{-0.03}^{+0.08}$ &$3.3 \pm 0.6$ &$0.011_{-0.001}^{+0.002}$ &$9.992_{-0.008}^{+0.004}$ &$50.5_{-5.3}^{+0.7}$ &$2.14 \pm 0.02$ &292(127)\\ 
1050020116 & $2.842_{-0.001}^{+0.084}$ &$1.002_{-0.002}^{+0.197}$ &$24.6_{-1.8}^{+1.3}$ &$0.05_{-0.005}^{+0.014}$ &$10.18_{-0.04}^{+0.11}$ &$4.1_{-0.4}^{+0.5}$ &$2.84 \pm 0.04$ &307(125)\\ 
1050020117 & $2.64_{-0.04}^{+0.08}$ &$1.00089_{-2e-05}^{+3e-05}$ &$19.8_{-0.2}^{+0.4}$ &$0.05 \pm 0.001$ &$10.048_{-0.004}^{+0.005}$ &$6.41_{-0.08}^{+0.04}$ &$2.82 \pm 0.01$ &218(121)\\ 
1050020118 & $2.57 \pm 0.01$ &$1.004_{-0.004}^{+0.024}$ &$290.1_{-0.8}^{+2.2}$ &$0.022 \pm 0.001$ &$11.14_{-0.03}^{+0.04}$ &$1.22_{-0.03}^{+0.02}$ &$2.64 \pm 0.01$ &208(118)\\ 
1050020119 & $3.0_{-0.2}^{+0.3}$ &$1.1 \pm 0.1$ &$14.0_{-2.9}^{+6.1}$ &$0.022_{-0.004}^{+0.001}$ &$10.014 \pm 0.002$ &$43.3 \pm 3.1$ &$2.5 \pm 0.1$ &265(126)\\ 
1050020121 & $2.589 \pm 0.001$ &$1.0001_{-0.0001}^{+0.0002}$ &$13.73_{-0.005}^{+0.004}$ &$0.04963 \pm 1e-05$ &$10.024 \pm 0.001$ &$8.908 \pm 0.002$ &$2.8113 \pm 0.0001$ &183(121)\\
\end{tabular}

\end{table}

\end{turnpage}
\begin{table}[h!]
\caption{\source Results of Reflection Modeling: Calculated Fluxes and Equivalent Widths}
\label{tab:refl_flux}
\tiny
\centering
\begin{tabular}{l|lllll|ll}
& \multicolumn{5}{c}{\sc Fluxes ($10^{-10}$ \fluxcgs)} \\
OBSID & Blackbody & Disk Blackbody & Powerlaw & \xillverco & Total & Line energy (keV) &EW (eV)\\

\hline \hline
1050020115 & $12.32 \pm 0.03$ &$12.32 \pm 0.01$ &$8.05 \pm 0.02$ &$11.42 \pm 0.03$ &$36.76 \pm 0.02$ &$0.684_{-0.003}^{+0.002}$ &$39.1 \pm 2.9$\\ 
1050020116 & $5.62 \pm 0.03$ &$5.57 \pm 0.01$ &$19.76 \pm 0.04$ &$13.95 \pm 0.02$ &$42.38 \pm 0.03$ &$0.688_{-0.003}^{+0.004}$ &$54.1 \pm 0.8$\\ 
1050020117 & $2.15 \pm 0.02$ &$3.01 \pm 0.01$ &$9.98 \pm 0.02$ &$10.16 \pm 0.02$ &$23.15 \pm 0.02$ &$0.6931_{-0.0003}^{+0.0104}$ &$15.8_{-7.6}^{+13.1}$\\ 
1050020118 & $1.73 \pm 0.02$ &$0.36 \pm 0.01$ &$13.58 \pm 0.02$ &$6.18 \pm 0.02$ &$20.68 \pm 0.02$ &$0.692_{-0.004}^{+0.002}$ &$58.5_{-0.8}^{+0.9}$\\ 
1050020119 & $9.15 \pm 0.02$ &$9.27 \pm 0.01$ &$3.8 \pm 0.02$ &$13.83 \pm 0.02$ &$29.57 \pm 0.02$ &$0.695_{-0.002}^{+0.003}$ &$52.7 \pm 0.8$\\ 
1050020121 & $2.15 \pm 0.02$ &$5.34 \pm 0.02$ &$6.56 \pm 0.04$ &$13.59 \pm 0.03$ &$25.81 \pm 0.03$ &$0.694_{-0.004}^{+0.003}$ &$53.8 \pm 1.1$\\
\end{tabular}

\end{table}

\setcounter{table}{0}
\renewcommand{\thetable}{C.\arabic{table}}
\section{\nustar Fit parameters}
This section of the appendix summarizes the results of reflection modeling for the \nustar spectra. Note: $\dagger$ indicates the parameter reached an upper bound and had no constraints on uncertainty. 
\begin{turnpage}
\begin{table}[h!]

\caption{\source Results of Reflection Modeling: \nustar}
\label{tab:nustar_refl}

\begin{tabular}{cc|llllll}
&  &30363002002  &30702009002 &30702009004 &30702009006 &30702009008 &30702009010\\
\hline
{\sc CONSTANT}&const &$0.968_{-0.003}^{+0.004}$ &$1.00\pm0.01$ &$0.996_{-0.004}^{+0.006}$ &$0.995_{-0.004}^{+0.006}$ &$0.996\pm0.001$ &$0.98\pm0.01$ \\
{\sc BBODY} & kT (keV) &$1.55_{-0.03}^{+0.02}$ &$1.16_{-0.04}^{+0.03}$ &$1.07_{-0.06}^{+0.03}$ &$0.65_{-0.04}^{+0.07}$ &$1.16_{-0.02}^{+0.01}$ &$1.06_{-0.03}^{+0.01}$\\
            & k$_{\mathrm{BB}}$ ($10^{-3}$) & $6.1_{-0.1}^{+0.4}$ &$2.2\pm0.3$ & $1.5_{-0.2}^{+0.1}$ &$1.50_{-0.38}^{+0.02}$ &$1.6\pm0.1$ & $1.46_{-0.19}^{+0.04}$ \\
{\sc DISKBB}& kT (keV) &$0.2\pm0.2$ &$0.50_{-0.04}^{+0.02}$ &$0.49_{-0.07}^{+0.01}$ &$1.7_{-0.5}^{+0.2}$ &$0.58_{-0.05}^{+0.02}$ &$0.54_{-0.02}^{+0.01}$ \\
            & k$_{\mathrm{DBB}}$  &$1117.4_{-1045.9}^{+1854.3}$ &$274.2_{-70.5}^{+118.7}$ &$388.8_{-73.8}^{+264.2}$ &$0.3_{-0.3}^{+0.5}$ &$100.3_{-27.7}^{+68.5}$ &$172.3_{-19.8}^{+19.3}$ \\
{\sc CUTOFFPL}& $\Gamma$  &$2.15_{-0.03}^{+0.06}$ &$2.15_{-0.08}^{+0.02}$ &$2.04_{-0.04}^{+0.07}$ & $1.84_{-0.01}^{+0.12}$ & $2.15_{-0.02}^{+0.01}$ &$2.06_{-0.01}^{+0.04}$ \\
            & E$_{\mathrm{cutoff}}$ (keV)  &$10.1_{-0.4}^{+0.5}$ &$100.7_{-22.0}^{+8.7}$ &$119.7_{-14.7}^{+30.6}$ &$85.9_{-4.1}^{+60.4}$ &$121.7_{-23.0}^{+12.7}$ & $199.0_{-27.7}^{+0.9}$\\
\end{tabular}

\end{table}

\end{turnpage}

\end{document}